Density Scaling and Decoupling in *o*-Terphenyl, Salol, and Dibutyphthalate

R. Casalini[1], S.S. Bair[2] and C.M. Roland[1]

[1]*Naval Research Laboratory, Chemistry Division, Code 6100, Washington DC 20375-5342*
[2]*Center for High Pressure Rheology, Georgia Institute of Technology, Atlanta, GA 30332-0405*

**Abstract:** We present new viscosity and equation of state (EoS) results extending to high pressures for *o*-terphenyl, salol, and dibutylphthalate. Using these and data from the literature, we show that the three liquids all conform to density scaling; that is, their reduce viscosities and reorientational relaxation times are a function of the ratio of temperature and density with the latter raised to a constant. Moreover, the functional form of the dependence on this ratio is independent of the experimental probe of the dynamics. This means that there is no decoupling of the viscosities and relaxation times over the measured range of conditions. Previous literature at odds with these results were based on erroneous extrapolations of the EoS or problematic diamond anvil viscosity data. Thus, there are no exceptions to the experimental fact that every non-associated liquid complies with density scaling with an invariant scaling exponent.

**Introduction**

Density scaling refers to the superpositioning of dynamic variables and transport coefficients when plotted as a function of the ratio of temperature, $T$, and density, $\rho$, with the latter raised to a constant [1]. For the viscosity, $\eta$, the scaling relation is

$$\eta = f\left(T\rho^{-\gamma}\right) \quad (1)$$

where $\gamma$ is a material constant and $f$ a function. The form of $f$ is usually unspecified, although an expression, which describes experimental data accurately [2,3], has been derived by assuming that the configurational entropy controls the dynamics [4]. There are two regimes for the response of $f$ to temperature and pressure [5]. For small $\rho^\gamma/T$, $\dfrac{d^2 \ln f}{d\left(\rho^\gamma/T\right)^2} < 0$ and for large values the second derivative is greater than zero. Over a sufficient range of density and temperature both regimes have been observed for the same material [6,7]. Density scaling is exact for a fluid whose intermolecular potential is a repulsive, inverse power law (IPL) [8,9], and molecular dynamic simulations (MDS) indicate it is a good approximation for materials with a more general interaction potential [10,11,12,13,14]. It applies generally to materials in which the non-bonded interactions are limited to van der Waals and Coulombic interactions, provided the latter are not sufficiently strong to form complexes. Hydrogen-bonded liquids deviate from eq. 1 because their chemical structure changes with $T$ and $\rho$ [15]. Although in principle no material (excepting hypothetical IPL fluids) conforms to density scaling for all state points [16], for simple liquids density scaling has been found to describe accurately the dynamics and transport properties over all experimentally accessible state points [1,17]. For example, recent measurements using a diamond anvil cell (DAC) have shown $T/\rho^\gamma$ scaling for cumene over densities changes of 28% [18] and nitrogen for $\Delta\rho$ that exceed 100% [19]. In some materials there does appear to be some deviation from eq. 1 for sufficiently large density changes; that is, $\gamma$ may not remain constant. However, such deviations entail long extrapolations of the equation of state (EoS) [20,21], which opens to question the accuracy of the results. Nevertheless, considering the failure of density scaling at very high $\rho$ in MDS [16], high pressure experimental data are of great interest.

If the density scaling exponent reflects of the steepness of the intermolecular potential [10,11,12,13,14], different dynamic quantities are expected to be described by the same scaling exponent. Thus, while different experiments, sensitive to different correlation functions, can yield relaxation times, $\tau$, having different magnitudes, the changes in $\tau$ due to changes in thermodynamic conditions should not depend on the experimental probe. Both the form of the scaling function $f$ and the exponent $\gamma$ should be material constants, independent of the measurement method. This expectation is borne out by results for many glass-forming materials [1,17], but there are a few apparent exceptions:

When various data for $o$-terphenol (OTP) are plotted versus $T/\rho^\gamma$, different scaling exponents are obtained; e.g., $\gamma$=4 for the viscosity [22] and dynamic light scattering relaxation times [23], but $\gamma$=4.25 for relaxation times measured by dielectric spectroscopy [24]. Salol presents similar problems. Although dielectric $\tau$ have been shown to conform to eq. 1 with $\gamma$ = 5.2 [25], a decoupling of $\tau$ and $\eta$ was observed [26]. If correct, this decoupling implies that $\tau$ and $\eta$ of salol cannot superpose for the same function of $T/\rho^\gamma$. For dibutylphthalate (DBP) there is no decoupling of the rotational relaxation times with either the viscosity [27] or translational diffusion of ions [28]. However, conformance to eq. 1 is uncertain, with different conclusions drawn from analyses of different data sets employing different EoS [20,29,30].

In this paper we address the apparent issues with OTP, salol, and DBP, showing that either an inadequate EoS or errors in published experimental results underlie the putative problems with density scaling. There is no experimental example of deviation from eq. 1 or of decoupling for these ordinary (non-associated) liquids.

**Experimental**

Viscosities were determined using three falling-cylinder viscometers [31]. In this method the vertical position of a magnetic cylindrical sinker with a tantalum core is monitored with an LVDT. The descent velocity depends on the geometry and the relative densities of the sinker and the liquid, with the viscosity of the latter obtained by calibration; the measurement error in $\eta$ is 4%. The pressure is measured to within 2 MPa by transducers calibrated against a manganin cell (Harwood Eng.). The viscosities of OTP and DBP were previously reported [32]; the viscosities for salol are newly determined.

To determine the EoS for OTP and DBP, volume changes were measured as a function of pressure and temperature with a Gnomix instrument [33]. A liquid (mercury) served as the confining medium, in order to maintain hydrostatic conditions when the sample solidifies by crystallization or vitrification. The differential data were converted to specific volumes, $V$ ($=\rho^{-1}$), using the value determined at ambient conditions by the buoyancy method (Archimedes' principle). At room temperature OTP is crystalline, so the PVT experiments included measurements on the solid, from which the absolute $\rho$ of the liquid could be obtained.

**Results**

**OTP.** A caveat in the application of density scaling is the requirement of an accurate EoS, in order to convert measured $P$-dependences to a dependence on $\rho$. For OTP this problem is exacerbated because the reported PVT data are limited to pressures less than 80 MPa [34]. To address this short-coming, we measured $V$ of OTP at pressures up to 200 MPa over temperatures for which the material is a liquid (see Figure S1 in Supplementary Material). The Tait equation is known to describe accurately PVT data for liquids [35]; fitting to the results for OTP we obtain

$$V = 0.9079\exp(7.546\times10^{-4}T)\left[1 - 0.2059\log\left(1 + \frac{P}{205.4\exp(-0.00458T)}\right)\right] \quad (2)$$

with $V$ in ml/g, $T$ in Celsius and $P$ in MPa. This EoS also adequately describes densities previously reported for the liquid [34,36,37], although it deviates from the prior EoS [34] at higher pressures.

In order to test eq. 1, we use eq. 2 to obtain $\rho$ for each measured state point. Strictly speaking, the scaling property applies to reduced quantities, defined as [9,38]

$$\eta^* = v^{2/3}\left(mRT\right)^{-1/2}\eta$$
$$\tau^* = v^{-1/3}\left(RT/m\right)^{1/2}\tau \quad (3)$$

in which $v$ is the molar volume, $m$ the molar mass, and $R$ the gas constant. The scaling plot of reduced variables is shown for OTP in Figure 1. To determine the best-fit value of the exponent, we use the fact that for a constant value of $\eta$ or $\tau$, a double logarithmic plot of $T^*$ versus $V^*$ has a slope equal to $-\gamma$. This power-law plot is shown in the inset of Fig. 1, yielding $\gamma = 5.35$ for OTP. (The isoviscous $T^*$ and $V^*$ values are tabulated in the Supplementary Material.) Such a procedure for obtaining the scaling exponent is objective and enables distinguishing between systematic deviations from eq. 1 and random scatter.

This value of $\gamma$ yields excellent superpositioning of both $\eta^*$ and $\tau^*$, the data collapsing to a single curve (same $f$); the ratio of $\eta/\tau$ is 0.3 GPa. Note that the scaling exponent required to superpose unreduced quantities is not necessarily the same [38]. The only data for OTP that do not superpose are those of Schug et al. [39]. However, the dielectric data of Naoki et al. for $P \leq 75$ MPa [40], the dynamic light scattering results of Fytas et al. [41], and the viscosity measurements herein for $P$ up to 403 MPa fall on a single curve, along with $\eta$ for ambient pressure [42,43,44].

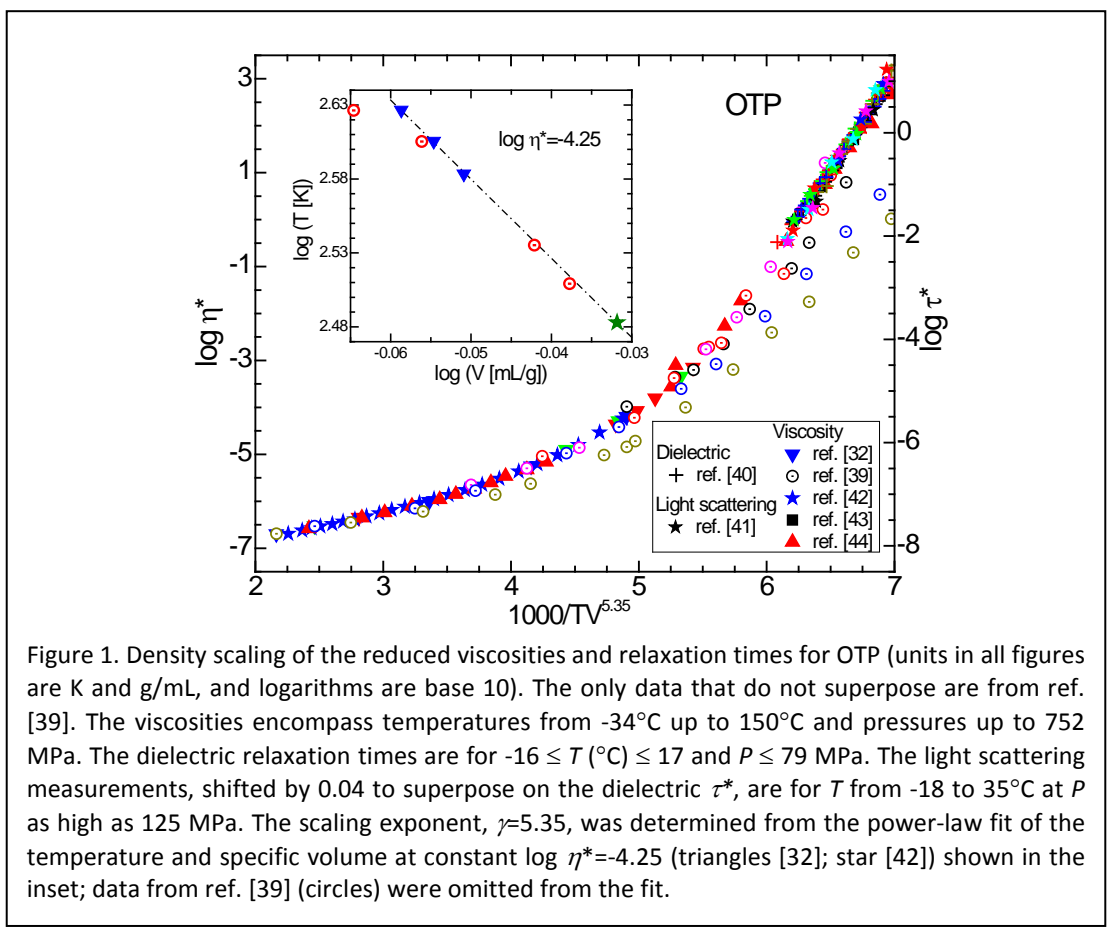

Figure 1. Density scaling of the reduced viscosities and relaxation times for OTP (units in all figures are K and g/mL, and logarithms are base 10). The only data that do not superpose are from ref. [39]. The viscosities encompass temperatures from -34°C up to 150°C and pressures up to 752 MPa. The dielectric relaxation times are for $-16 \leq T$ (°C) $\leq 17$ and $P \leq 79$ MPa. The light scattering measurements, shifted by 0.04 to superpose on the dielectric $\tau^*$, are for $T$ from -18 to 35°C at $P$ as high as 125 MPa. The scaling exponent, $\gamma=5.35$, was determined from the power-law fit of the temperature and specific volume at constant log $\eta^*=-4.25$ (triangles [32]; star [42]) shown in the inset; data from ref. [39] (circles) were omitted from the fit.

**Salol.** To reconcile the discrepancy between the density scaling of dielectric relaxation times [25] but the purported decoupling of $\tau$ and $\eta$ [26], we measured the viscosity of salol at three temperatures for pressures up to 0.4 GPa. These data are reproduced in Table S2 in the Supplementary Materials. As seen in Figure 2, when plotted versus $T/\rho^{5.2}$ (with the scaling exponent again obtained from a power-law plot of $T^*$ vs. $V^*$; see Table S1 in the Supplementary Materials), the $\eta^*$ collapse to a single curve, along with the $\tau^*$ from dielectric spectroscopy [25] and dynamic light scattering [45]. The ratio of $\eta/\tau$ is 0.1 GPa. As was the case for OTP, the only departures from eq. 1 are the viscosities from Schug et al. [39]. These are the same data that led to the previous conclusion, not borne out by Fig. 2, that $\eta$ and $\tau$ for salol are decoupled [26].

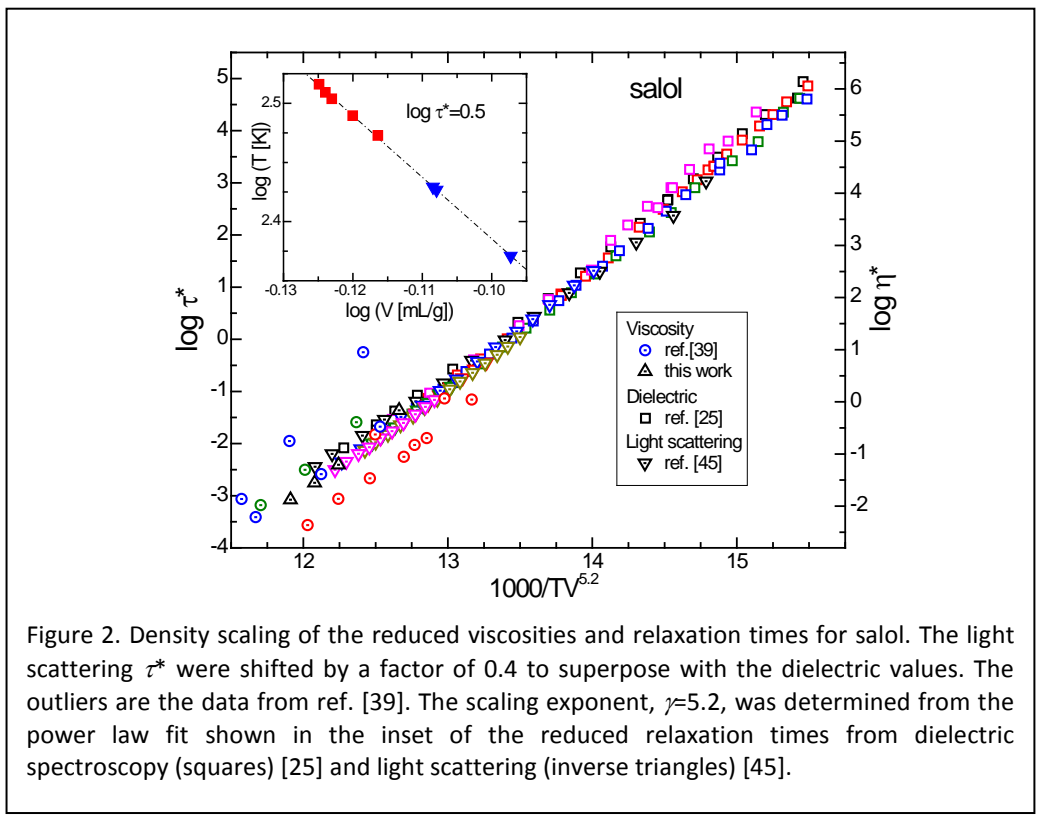

Figure 2. Density scaling of the reduced viscosities and relaxation times for salol. The light scattering $\tau^*$ were shifted by a factor of 0.4 to superpose with the dielectric values. The outliers are the data from ref. [39]. The scaling exponent, $\gamma$=5.2, was determined from the power law fit shown in the inset of the reduced relaxation times from dielectric spectroscopy (squares) [25] and light scattering (inverse triangles) [45].

**Dibutylphthalate.** Given the apparent errors in the high pressure viscosities for OTP and salol in ref. [39], we revisited the glass-forming liquid DBP, whose conformance to density scaling has been of interest to several groups. Casalini and Roland [29] measured the EoS for DBP, and found that the dielectric relaxation times in ref. [28] were a function of $T/\rho^{3.2}$. Niss et al. [30], using older PVT data of Bridgman [46], observed conformance with eq. 1 and $\gamma$=2.5 for several sets of $\tau$ data; however, there was a conspicuous departure of the DAC viscosity data from Cook et al. [47]. This is the same group [39] that reported the discordant viscosities in Figs. 1 and 2 for OTP and salol, respectively. Subsequently, Bøhling et al. [20] measured higher pressure $\tau$ data for DBP and found departures from density scaling; this analysis also relied on the EoS from older PVT data [46].

There is no reason to expect a breakdown of density scaling for a non-associated liquid like DBP. The suspicion is that deviations from eq. 1 may be a consequence of using a faulty EoS obtained by extrapolating PVT measurements [29,46]. Accordingly, we carried out new PVT measurements, combined with the data from ref. [46], to obtain the following for the Tait equation

$$V = 0.9396 \exp(7.775 \times 10^{-4} T) \left[ 1 - 0.2252 \log\left( 1 + \frac{P}{203.4 \exp(-0.00465T)} \right) \right] \quad (4)$$

Applying this to the available dielectric relaxation times [28,30] and viscosities, both from the literature [47,48] and measured previously to 1.25 GPa [6], we obtained the scaling plot in Figure 3. There is fair collapse of the reduced quantities; the ratio of $\eta/\tau$ = 0.1 GPa. The marked departures are two $\eta$ values at the highest pressures measured by Cook et al. [47]. The common scaling of the dynamic variable for the same $\gamma$=2.94 is consistent with the absence of decoupling for this material [27] deduced from the translational diffusion behavior of ions [28].

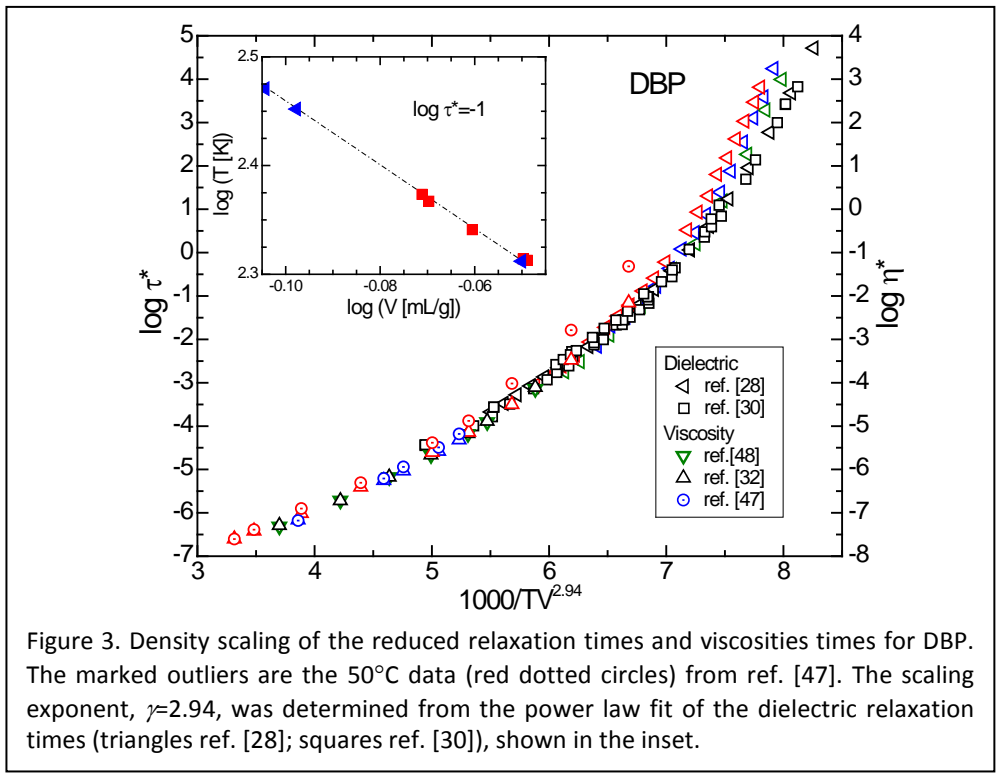

Figure 3. Density scaling of the reduced relaxation times and viscosities times for DBP. The marked outliers are the 50°C data (red dotted circles) from ref. [47]. The scaling exponent, $\gamma$=2.94, was determined from the power law fit of the dielectric relaxation times (triangles ref. [28]; squares ref. [30]), shown in the inset.

**Summary**

Density scaling and the coupling of rotational and translational motions are important properties both fundamentally, toward understanding the glass transition problem, and for applications. The latter extend broadly, two examples being the nature of the materials in the earth's core and lower mantle [49,50], and lubricants and the mechanical energy loss in machines [51]. One of the appeals of density scaling is its generality, which makes ostensible exceptions to the behavior are significant.

In this work we show that density scaling is valid for OTP, salol, and DBP, with a common function (eq. 1) and exponent $\gamma$ describing the viscosity and relaxation times (in reduced units) for each material. This means there is no decoupling of $\eta$ and $\tau$. Previous indications of deviations from superpositioning of these quantities versus $T/\rho^\gamma$ or of different $T$- or $P$-dependences were based on inaccurate EoS and erroneous DAC measurements of $\eta$. Unfortunately, due to the scarcity of viscosity data at elevated pressures, extensive use has been made of these early DAC results [39,47]. The work was pioneering in the characterization of $\eta$ at very high pressures; however, as we show herein for OTP, salol, and DBP, the $\eta(P)$ are not accurate, probably due to systematic errors in the pressure. The cause of the putative errors is unknown. Pressures were determined in refs. [39,47] from the shift of the ruby fluorescence. This method is common, but known to have potential errors, for example from stress anisotropy [52] or the chromium content of the ruby affecting the latter's temperature dependence [53]. The salient point is that conclusions based on the earlier high pressure viscosity data [39,47] require reexamination.

**Supplementary Material**

See supplementary material for PVT data for OTP and DBP, the new viscosity measurements on salol, and the isoviscous/isochronal results plotted in the figure insets.

**Acknowledgement**

The work at NRL was supported by the Office of Naval Research. SSB was supported by the Center for Compact and Efficient Fluid Power, a National Science Foundation Engineering Research Center funded under cooperative agreement number EEC-0540834.